\documentclass[twoside,12pt]{article}
\usepackage{epsfig}

\newcommand{\be}{\begin{equation}}
\newcommand{\ee}{\end{equation}}
\newcommand{\bea}{\begin{eqnarray}}
\newcommand{\eea}{\end{eqnarray}}

\begin{document}

\title{ \vspace{1cm} Electromagnetic interactions at RHIC and LHC}
\author{M. Cem G\"{u}\c{c}l\"{u}
\\
Istanbul Technical University,Physics Department, Maslak-Istanbul, TURKEY}
\maketitle
\begin{abstract} At LHC energies the Lorentz factor will be 3400 for the Pb + Pb 
collisions and the electromagnetic interactions will play important roles. Cross
sections for the electromagnetic particle productions are very large and can not
be ignored for the lifetimes of the beams and background.  In this article, we are 
going to study some of the electromagnetic processes at RHIC and LHC and show the 
cross section calculations of the electron-positron pair production with the giant dipole 
resonance of the ions.
\end{abstract}
\section{Introduction}

For the LHC and RHIC energies, cross section of the lepton pair production is 
very large for the Pb + Pb and Au + Au collisions. In this work, we focus mainly on ultra-peripheral 
collisions (UPC) of heavy ions for $b > 2R$ region. In this type of collisions, purely 
electromagnetic interactions and photo-nuclear interactions are the dominant 
processes ~\cite{hbt04,ja04} . Production of free $e^{+}e^{-}$ pairs and bound-free 
production of $e^{+}e^{-}$ pairs play important roles at RHIC and at LHC. Free 
electron-positron pair production cross sections are about 36 kbarn at RHIC 
energies (Au+Au collisions) and about 227 kbarn at LHC energies (Pb+Pb collisions). 
On the other hand bound-free pair production cross sections are about 80 barns at RHIC 
energies and about 200 barns at LHC energies for the corresponding ion collisions. 
Therefore it is an important contribution to the background and with the electromagnetic 
excitation of the giant dipole resonance play important role for limitation to the lifetime 
of the beams. All these electromagnetic processes result in a change in the charge of the 
ions in the beams. 

In Table 1. we have tabulated the parameters for accelerators SPS, RHIC and LHC. $E_{crit}$ 
is the critical electric field, and $E_{max}$ is the maximum electric field to produce 
electron-positron pairs. As we clearly see that for higher energies the ratio of maximum electric field to
critical electric field is become larger. 

\begin{table}
\begin{center}
\begin{minipage}[t]{16.5 cm}
\caption{\label{tab:table1}Integrated cross sections for Au-Au collisions at RHIC energies for STAR 
experimental restrictions. First row shows the calculations of Hencken et al. Second row is obtained 
by Monte Carlo QED calculation and third row shows the same calculation with Woods-Saxon
nuclear form-factor.}
\begin{tabular}{cccc}
\\
\hline
Parameters&SPS&RHIC&LHC \\
\hline
$\gamma$ & 10& 100& 3400 \\
$E_{max}$(V/m)  & $10^{20}$ & $10^{22}$ & $10^{26}$ \\
$E_{max}/E_{crit}$ & 1.6 & 160 & 54,000 \\
\end{tabular}
\end{minipage}
\end{center}
\end{table}
There are many attempts to calculate electromagnetic lepton pair production cross section. 
The two-photon process ~\cite{W34,bkt,bb,BS89} has been modeled through the equivalent-photon 
approximation. In this model, the equivalent-photon flux associated with a relativistic charged 
particle is obtained via a Fourier decomposition of the electromagnetic interaction. Cross 
sections are obtained by folding the elementary, real two-photon cross section for the pair 
production process with the equivalent-photon flux produced by each ion. Although the results 
for the total cross sections are reasonably accurate, however, the details of the differential 
cross sections, spectra, and impact-parameter dependence differ. The method loses 
applicability at impact parameters less than the Compton wavelength of the lepton, which 
is the region of greatest interest for the study of nonperturbative effects. 

Most of the calculations show that cross sections of free pairs production agree with
the following equation
\be
\sigma_{free\; pairs} \sim Z_{1}^{2} Z_{2}^{2} ln(\gamma)^{3} \; . \label{eq:tmat}
\ee
where $Z_{1}$ and $Z_{2}$ are the charge numbers of the target and projectile nuclei, and
$\gamma$ is the Lorentz factor. We have also calculated free lepton pair production numerically
by using the Monte-Carlo techniques and found out that our results agree with the above equation.
Free pair production cross section is proportional to the energy and square of the charges of the 
colliding nuclei.

When an electron-positron pair is produced,
one of the electron can be bound to one of the incident nuclei. This electromagnetic
processes is called bound-free pair production (it is shown in figure 1.) and the cross section is about 280 b
for Pb + Pb collisions at LHC and is about 80 b for Au + Au collisions at RHIC energies.
The bound-free pair production cross section plays important roles for the monitoring 
luminosity of the heavy-ion beams at the LHC.  
There are many different calculations for capture into a K-shell orbit,
\be
\sigma_{capture} \sim Z_{1}^{5} Z_{2}^{2} ln(\frac{\gamma}{\Delta}) \; . \label{eq:tmat}
\ee
where$ Z_{1}$ is the charge number of the ion capturing electron, and $\Delta$ is a
slowly varying parameter.  Recent report ~\cite{bjg} shows the first measurement of the beam losses
due to electron capture at RHIC. 100 GeV/nucleon $^{63}Cu^{29+}$ ions are used at RHIC, and
the measurement confirms the order of magnitude of the theoretical calculations.

We have obtained cross section expressions for  electron-positron pair production from relativistic 
heavy ion collision based on a lowest order QED calculation. Our Lagrangian consists of three terms:
\begin{eqnarray}
\mathcal{L}_{QED} & = & \mathcal{L}_{e^{+}e^{-}} + \mathcal{L}_{EM} + \mathcal{L}_{int} \nonumber \\
 & = & \bar{\Psi}(i\hat{\partial} - m_{e})\Psi - \frac{1}{4}F_{\mu \nu}F^{\mu \nu} - e\bar{\Psi}\gamma^{\mu}\Psi A_{\mu} 
\end{eqnarray}
Here $\mathcal{L}_{e^{+}e^{-}}$ is the term for the free electrons-positrons, $\mathcal{L}_{EM}$ is 
the electromagnetic field and $ \mathcal{L}_{int}$ is the interaction term between the leptons.
At RHIC and LHC energies, for small  impact parameters pair production probabilities violates the unitarity. 
Therefore lowest order diagrams do not describe the pair production process and higher order terms must be included

Because of the some technical difficulties, we have not fully tested the theoretical calculations and experimental 
results of electron-positron pair production. Vane at al. ~\cite{vane} has obtained experimental data at SPS for 
lepton pair production. In comparison with data, we find that the two-photon external-field model does quite well 
for the region where a majority of the pairs are being produced. However, for the high-energy (or high-momentum ) 
tail, the theory is under predicting the data.

In this work, we calculate the probability of electron-positron pair production in lowest order QED. 
We use semi-classical approximation in the calculation and use Monte Carlo method to obtain exact 
results. We then compare our results with the STAR Collaboration and other theoretical calculations.

\section{Calculation of $e^{-}$ $e^{+}$ pair production with giant dipole resonance}
Recently, the STAR Collaboration has measured electron-positron pairs together with the electromagnetic excitation 
of both ions, predominantly to the giant dipole resonance. In such measurement, it is assumed that no hadronic 
interactions occur and minimum impact parameter is twice the nuclear radius. The STAR Collaboration used Gold 
atoms at $\sqrt{s_{NN}}$ = 200 GeV per nucleon energies. The decay of the excited nucleus generally emits one 
or two neutrons and these neutrons are detected in the forward Zero Degree Calorimeter (ZDC).

\begin{figure}[tb]
\begin{center}
\begin{minipage}[t]{8 cm}
\epsfig{file=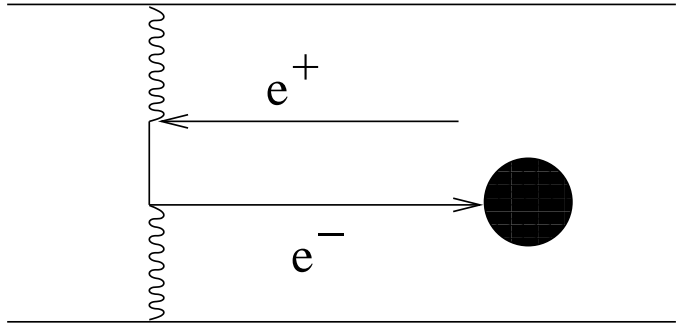,scale=0.7}
\end{minipage}
\begin{minipage}[t]{16.5 cm}
\caption{Diagram for pair production with capture (bound-free pair production) in relativistic heavy-ion
collision.}
\end{minipage}
\end{center}
\end{figure}
The STAR detector measures the produced electron-positron pairs for the limited kinematic range 
of pair mass 140 MeV $<$ $M_{ee}$ $<$ 265 MeV, pair rapidity $ |Y|$ $<$ 1.15 and the transverse 
momentum $p_{\perp}$ $>$ 65 MeV. If the pair production is independent of the nuclear excitation, 
the total cross section of electron-positron pair production with Giant Dipole Resonance can be written as
\begin{equation}
\sigma^{GDR}_{e^{-}e^{+}}=2\pi\int_{\rho_{min}}^{\infty} d\rho \rho P_{e^{-}e^{+}}(\rho)P_{GDR}^{2}(\rho)
\end{equation}

where $P_{e^{-}e^{+}}$ is the probability of electron-positron pair production and $P_{GDR}(b)$ is 
the probability of a simultaneous nuclear excitation as a function of impact parameter.
\begin{figure}[tb]
\begin{center}
\begin{minipage}[t]{8 cm}
\epsfig{file=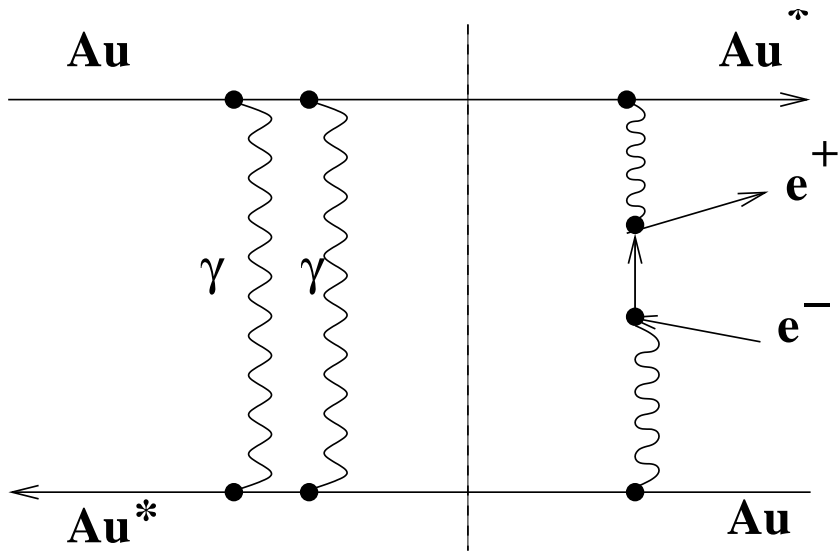,scale=0.7}
\end{minipage}
\begin{minipage}[t]{16.5 cm}
\caption{Electron-positron production (on the right) with a mutual Coulomb excitation 
(on the left) mainly giant dipole resonance (GDR).
These two processes are independent from each other.}
\end{minipage}
\end{center}
\end{figure}
\begin{figure}[tb]
\begin{center}
\begin{minipage}[t]{8 cm}
\epsfig{file=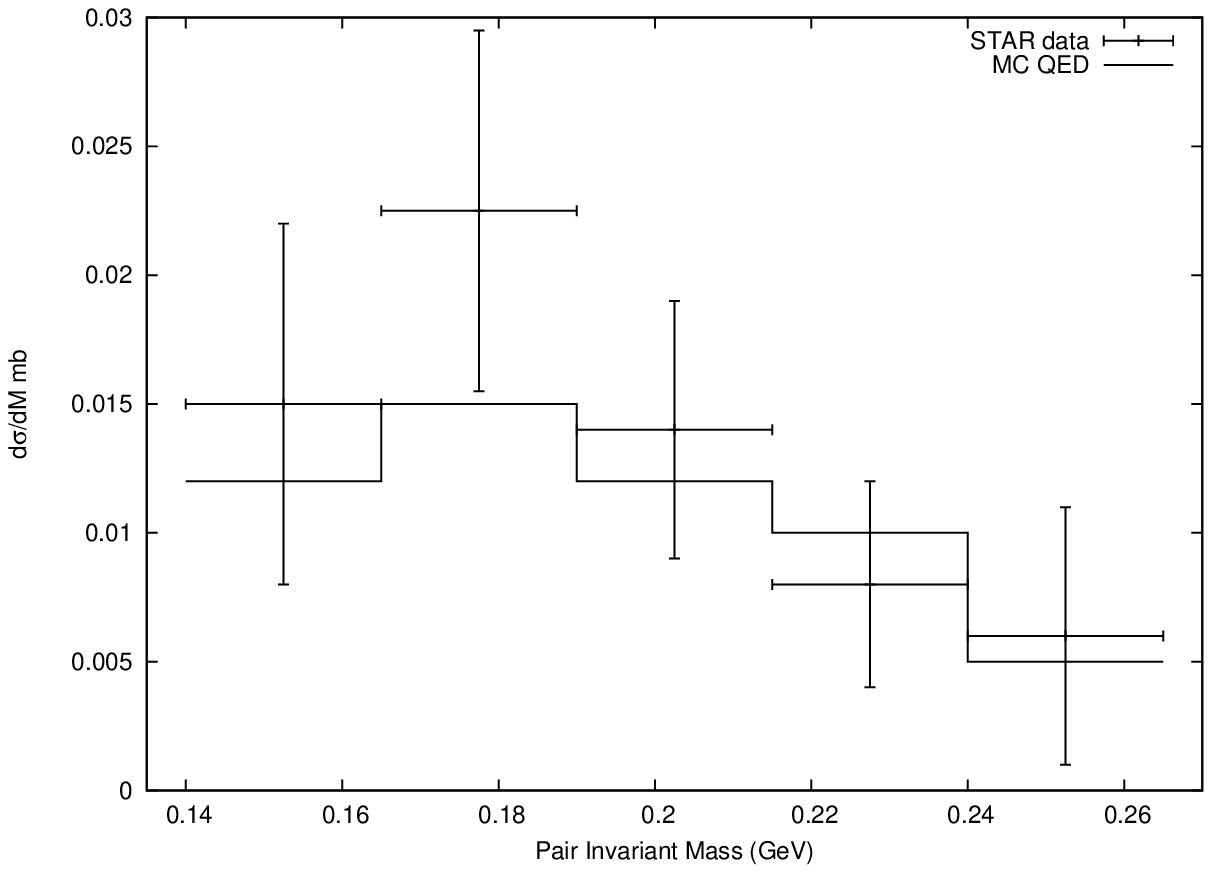,scale=0.9}
\end{minipage}
\begin{minipage}[t]{16.5 cm}
\caption{Differential cross section as a function of invariant pair mass, within the STAR
kinematic restrictions. }
\end{minipage}
\end{center}
\end{figure}
Previously, we have calculated lepton-pair production by using the lowest order Feynman 
diagrams. According to this QED calculation we have obtained the total pair cross section as

\begin{eqnarray}
P_{e^{-}e^{+}}(\rho) = \frac{1}{2\pi\rho}\frac{d\sigma}{d\rho} =\frac{1}{2\pi} \sigma_{T}\frac{\rho_{0}}{(\rho_{0}^{2}+\rho^{2})^{3/2}}
\label{dsdb}
\end{eqnarray}
here $\sigma_{T}$ is the total cross section of free lepton pair production cross section and it is
equal to $\sigma_{T} = Z_{1}^{2} Z_{2}^{2} ln(\gamma)^{3}$ and $\rho_{0}$ is the parameter that is
obtained numerically. We have calculated the above function for Au + Au collisions 
for $\gamma$ = 10, 100 and 3400. This particular calculations show that for a fixed values 
of one variable in the equation is in exponential form. By examining the fluctuations between 
the points for different values of a fixed variable, and from standard estimates for the errors 
of a Monte Carlo integration, we take sufficient points until we believe that each points has 
converged to within five percent. A smooth function is then fit to the calculated numbers. 

We have determined the constants $\rho_{0}$ at RHIC energies for STAR parameters and without 
any restrictions. It is equal to $1.35 \lambda_{C}$ without any restrictions and $0.114 \lambda_{C}$ 
for the STAR kinematic restrictions. When we integrate this impact parameter dependence cross 
section over the impact parameter, we obtain the total cross section, and the 
parameter $\rho_{0}$ is disappear in the total cross section. In this calculation, there is no kinematic 
restriction for pair mass, rapidity and transverse momentum. In order to compare with the STAR 
experiment, we also restrict the pair rapidity $ |Y|$ $<$ 1.15 and the transverse 
momentum $p_{\perp}$ $>$ 65 MeV. The total untagged cross section is 0.322 barn 
and it is in perfect agreement with Kai Hencken et al. calculations.

\begin{figure}[tb]
\begin{center}
\begin{minipage}[t]{8 cm}
\epsfig{file=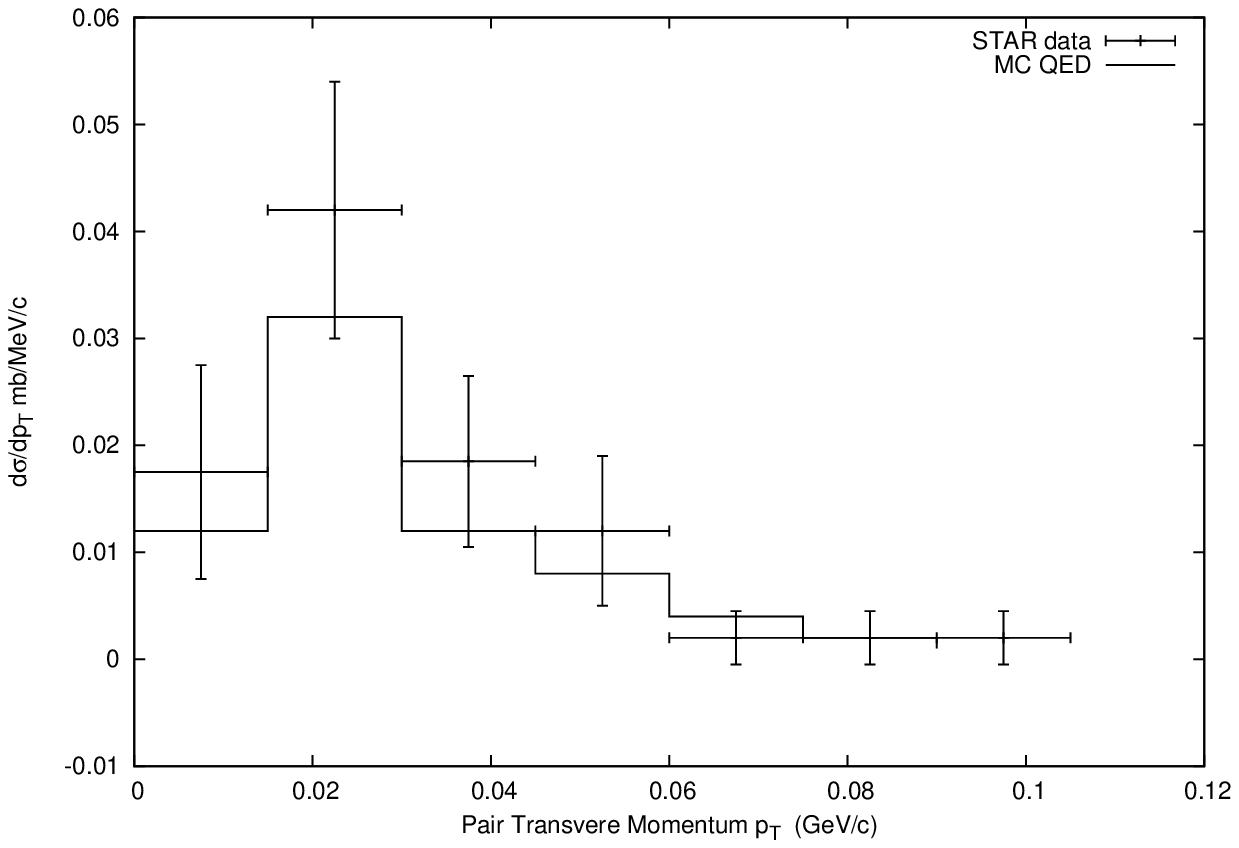,scale=0.9}
\end{minipage}
\begin{minipage}[t]{16.5 cm}
\caption{Differential cross section as a function of transverse momentum of the produced electrons, 
within the STAR kinematic restrictions.}
\end{minipage}
\end{center}
\end{figure}

Recent paper ~\cite{ja04,srk2,npa03} reports on electromagnetic production of electron-positron pair 
accompanied by giant dipole resonance (GDR) at RHIC for Au-Au collisions. This process 
is shown in figure 2 and in addition to the lepton pairs, the nuclei exchange also some 
photons and this may break up the nuclei.
For the probability of GDR excitation in one ion we use the approximation
\begin{eqnarray}
P_{GDR}(\rho) = S/\rho^{2}
\end{eqnarray}
here 
\begin{eqnarray}
S = \frac{2\alpha^{2}Z^{3}N}{Am_{N}\omega} = 5.45 \times 10^{-5}Z^{3}N A^{-2/3}fm^{2}
\end{eqnarray}
where $m_{N}$ is the nucleon mass, $N$ is the neutron number, $Z$ is the proton number 
and $A$ is the mass number of the ions, respectively. More information can be found in  
Ref. ~\cite{hbt04,bha03,vbm}. We can calculate the total cross section for the electron-positron 
pair production accompanied by nuclear dissociation in peripheral heavy-ion collisions.
Since for small values of impact parameter $\rho_{min}$, the probability for GDR excitation 
is large, we include multi-photon excitation probability as
\begin{eqnarray}
P(\rho)= 1 - exp[-P_{GDR}(\rho)].
\end{eqnarray}
This equation represent the multi-photon excitation probability and restore the unitarity. 

\section{Results}

In our Monte Carlo calculation, we have set the integral limits for the STAR experimental restrictions. 
We have increased the random numbers until the results are converged. We have used 10 millions 
Monte Carlo points for each variables. The error is less than 5 percent.
We have calculated the integrated cross section for Au-Au collisions at RHIC energies with the 
restrictions  $p_{\perp}$ $>$ 65 MeV/c, $ |Y|$ $<$ 1.15. Hencken $et\; al.$ results are 
$2.30$ mb, $1.76$ mb and $1.43$ mb for the the minimum  impact parameters $\rho_{min}=13, 14, 15$ fm
respectively. For the corresponding minimum impact parameters, we have obtained the tagged cross sections 
by Monte Carlo QED calculations and our results are $1.98$ mb, $1.50$ mb and $1.33$ mb respectively.
We have also used the same calculation with Woods-Saxon nuclear form factor. 
Although our results are smaller than the Hencken et al. calculations, the 
overall agreement is good. On the other hand, Baltz ~\cite{baltz, srk} has calculated total
cross section including the higher order QED effect and obtained a result of $1.67$ mb, and it
is an excellent agreement with the STAR data which is 1.65 $\pm$ 0.23 (stat) $\pm$ 0.30(syst) mb.
This could be an indication that higher-order QED effects play important role in lepton
pair production in RHIC and LHC.

In Figure 3, we have compare our pair-mass distribution with the STAR experiment. 
The solid line shows our calculation. Our distribution is in the range of the error bars. 
For pair mass around 180 GeV, our result slightly underestimates the measured values. 
In Figure 4, we have also shown our calculation of the pair transverse momentum distribution. 
This distribution is also in the range of the error bars. Baltz has calculated both invariant
mass distribution and pair transverse momentum distribution. He has included the higher
order effects in his calculation and the result is significantly lower than the simple QED results.

\section{Conclusion}
We have calculated the electron-positron pair production cross section exactly by using the 
Monte Carlo method. In our calculation, although we have not included the higher order effects, 
our  results are in good agreement with the Baltz calculation. Both calculations are also
in good agreement with the experimental results. The calculation done by Hencken at al
are also in good agreement with the experimental results, however higher than our calculations.

In future works, we are planning to include the higher order effects in our calculations.
This may also reduce the lepton pair production cross sections and the relevant distributions.
It is important to understand the contribution of the higher-order QED effects in 
relativistic heavy ion collisions. RHIC produces  the highest available electromagnetic fields from the
heavy ions. Unfortunately, the test of QED at strong fields is not the main goal of RHIC. 
We need further experiments  to be done at RHIC and also at LHC to answer these questions. 
In future experiments at RHIC and at LHC, there are some plans to improve the experimental 
conditions to measure the higher order effects in lepton pair productions. We may also 
understand the higher order Coulomb effects in these collisions.

O the other hand, GSI at Darmstadt has planned to build a new accelerator facility 
FAIR (Facility for Anti-proton and Ion Research ) ~\cite{hhg} with anti-proton and ion beams. 
Different kind of physics experiments will be performed at FAIR. One of the 
main goal of the FAIR project is to study " Quantum Electrodynamics QED at strong field".
 Although the above equations 1 and 2 are valid for the ultra-relativistic
energies, when the FAIR project is completed in 2016, the available Lorentz factor $\gamma$ will be up to
20 in fixed-target experiments. This could help us do experiments at intermediate relativistic energies.
Studying the differential cross sections as a function of impact parameter, and bound-free pair 
production at this intermediate relativistic energies are some of the main goals of the FAIR.
  
\section{Acknowledgments}                                                                  
This research is partially supported by the Istanbul Technical University.
I personally thank to S. R. Klein for valuable advise in calculating the cross sections.
I would like to thank also to Amand Faessler and Jochen Wambach  for supporting me
to participate to this workshop. Finally, I would like to thank to Hans Gutbrod for inviting
me to GSI to study the FAIR project, and I would also thank to DAAD (Deutscher Akademischer 
Austausch Dienst) for supporting me on this visit.

\end{document}